\documentclass[%
 reprint,
 superscriptaddress,
 nobibnotes,
 amsmath,amssymb,
 prappl,
]{revtex4-2}

\usepackage{hyperref}
\usepackage{graphicx}
\usepackage{dcolumn}
\usepackage{bm}
\usepackage{multirow}
\usepackage{booktabs}
\usepackage{rotfloat}
\usepackage{esint}
\usepackage{mathtools}

\usepackage{psfrag,color}
\usepackage{lipsum}

\usepackage{xcolor}
\usepackage[export]{adjustbox}

\usepackage{cellspace}
\usepackage{scalerel}

\newcommand{\m}{\mathrm{m}}

\newcommand{\pati}[1]{}

\makeatother

\begin{document}
\preprint{APS/123-QED}

\title{Scattering at Interluminal Interfaces}


\author{Zhiyu Li}
\email{lizhiyu@stu.xjtu.edu.cn}%
\affiliation{
Department of Electrical Engineering, Xi’an Jiaotong University, Xi’an, Shaanxi, China
}%

\author{Klaas De Kinder}
\affiliation{
Department of Electrical Engineering, KU Leuven, Leuven, Belgium
}%

\author{Xikui Ma}
\affiliation{
Department of Electrical Engineering, Xi’an Jiaotong University, Xi’an, Shaanxi, China
}%

\author{Christophe Caloz}
\affiliation{
Department of Electrical Engineering, KU Leuven, Leuven, Belgium
}%

\date{\today}

\begin{abstract}
Scattering at interluminal modulation interfaces, where a sharp space-time perturbation moves at a velocity lying between the wave velocities of the two surrounding media, has remained an open problem for decades.
This regime is somewhat reminiscent of the Cherenkov regime, in which the velocity of a charged particle exceeds the phase velocity of light in a medium. However, because it involves two media and a moving interface, it gives rise to richer and more complex scattering dynamics, with a single scattered wave when the incident wave propagates in the same direction as the interface and three scattered waves when they propagate in opposite directions.
Existing studies address only limited non-magnetic configurations, and a general formulation has yet to be established.
In this paper, we present a complete and general solution to scattering in the interluminal regime using a symmetric decomposition approach based on subluminal and superluminal limit interfaces, together with a space-time impulse response.
This approach provides clear physical insight into the scattering features of the interluminal regime. Our results bridge the long-standing gap between the subluminal and superluminal regimes and elucidate the fundamental mechanisms underlying interluminal scattering.
\end{abstract}

\maketitle

\section{Introduction}
\pati{STEMs Background \& Applications}
Space-time modulation systems~\cite{Biancalana_2007_dynamics,Caloz_2019_spacetime1,Caloz_2019_spacetime2,Yin_2022_floquet,Caloz_GSTEMs}, characterized by external modulation of material parameters in both space and time, have recently opened new possibilities in microwave and optical technologies. By breaking space-time symmetry and, consequently, conventional energy-momentum conservation laws, these systems enable a variety of novel applications, including magnetless nonreciprocity~\cite{Estep_2014_nonreciprocity,Correas_2016_NonreciGraphene,Taravati_2017_nonreciprocal}, parametric amplification~\cite{Tien_1958_parametric,Reed_2003_color,Galiffi_2019_broadband,Pendry_2021_gain}, homogenized dynamic media~\cite{Huidobro_2021_homogenization,Serra_2023_homogenization} and modulation-induced wave dragging~\cite{Huidobro_2019_fresnel,Prudencio_2023_replicating}.
Among them, \emph{modulation-interface systems}, in which the transition width of the medium discontinuity is much smaller than the wavelength, act as the fundamental building blocks~\cite{Caloz_GSTEMs} or ``space-time atoms'', and play a key role in classical and quantum wave manipulation~\cite{Caloz_2025_structuring}.

\begin{figure}[ht!]
    \centering
    \includegraphics[width=8.6cm]{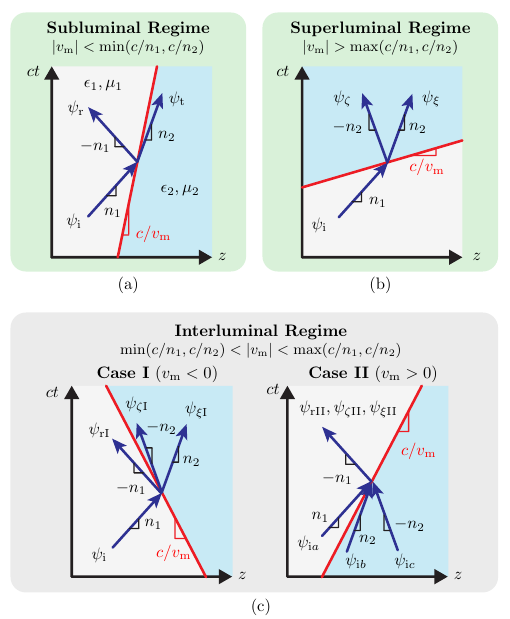}
    \caption{\label{fig:regimes} Wave scattering at different modulation interfaces. (a)~Subluminal regime, where the interface moves slower than the wave. (b)~Superluminal regime, where the interface moves faster than the wave. (c)~Interluminal regime, where the interface moves at a velocity between the wave velocities. The right panel illustrates three incident cases, each generating a single scattered wave: reflected for $\psi_{\mathrm{i}a}$, later-backward for $\psi_{\mathrm{i}b}$ and later-forward for $\psi_{\mathrm{i}c}$. The subscripts `r', `t', `$\zeta$' and `$\xi$' denote the reflected, transmitted, later-backward and later-forward waves, respectively; `I' and `II' indicate the scattered waves in Cases I and II.}
\end{figure}

\pati{Different Regimes \& Interluminal Interfaces}
Modulation-interface systems can be divided into three categories depending on the modulation velocity regime. In the subluminal regime [Fig.~\ref{fig:regimes}(a)], the interface moves slower than the wave, i.e., $|v_{\m}|<\min(c/n_1,c/n_2)$ with $n_{1,2}=\sqrt{\epsilon_{\mathrm{r}1,2}\mu_{\mathrm{r}1,2}}$ denoting the refractive indices of the media on either side of the interface, producing space-like scattering, with reflected and transmitted scattered waves. In the superluminal regime [Fig.~\ref{fig:regimes}(b)], the interface moves faster than the wave, i.e., $|v_{\m}|>\max(c/n_1,c/n_2)$, producing time-like scattering, with later-backward and later-forward scattered waves.
In the interluminal regime [Fig.~\ref{fig:regimes}(c)], which remains largely unexplored to date, the interface velocity lies in between the two wave velocities, i.e.,
$\min(c/n_1,c/n_2)<|v_{\m}|<\max(c/n_1,c/n_2)$.
This regime was historically referred to as the ``sonic regime''~\cite{Oliner_1961_wave,Cassedy_1962_temporal,Cassedy_1963_dispersion1} as the electromagnetic wave behavior in it is analogous to that of acoustic waves in a fluid at velocities between the subsonic and supersonic limits~\cite{Shui_2014_one}.
In electromagnetics, this regime can be related to the Cherenkov regime~\cite{Cerenkov_1934_visible,Kong_2008_emwtheory}, in which a charged particle radiates when it propagates faster than the velocity of light in the medium, $c/n$. However, modulation-interface and Cherenkov systems differ fundamentally. The former involves two media rather than one, subsequently corresponds to a finite velocity interval rather than a single velocity threshold and concerns the scattering of an electromagnetic wave rather than the radiation emitted by a charged particle.

\pati{Gap}
Unlike the well-established scattering problems in the subluminal and superluminal regimes, the interluminal regime represents a long-standing challenge. Early investigations by Ostrovskiĭ and Solomin~\cite{Ostrovskii_1967_correct,Ostrovskii_1975_some} provided partial insight, and more recent studies have revisited the problem~\cite{Biancalana_2007_dynamics,Deck_2019_well,Deck_2019_scattering}. However, existing work is restricted to non-magnetic media, and a complete and general analytical solution is still missing. The lack of a such a solution not only limits our understanding of wave dynamics in the interluminal regime but also leaves an inapplicable gap between the subluminal and superluminal regimes.

\pati{Contributions}
In this paper, we propose a symmetric decomposition approach for the interluminal scattering problem that is applicable in all generality. The method adapts the acoustic weak solution in~\cite{Shui_2014_one} to electromagnetic systems and generalizes it to arbitrary temporal wave profiles using the impulse response approach described in~\cite{Li_2025_Chirp}. We further provide new insights into velocity-independent scattering coefficients and the formation of shock waves. The validity of the approach is demonstrated through both analytical derivations and full-wave simulations.

\section{Recall of the Interluminality Problem}

Figure~\ref{fig:regimes}(c) illustrates the wave scattering phenomenology at different interluminal interfaces. For simplicity, we denote the rarer medium as medium~1 and the denser medium as medium~2 throughout the paper. In the contramoving case, where the interface moves backward [left panel of Fig.~\ref{fig:regimes}(c)], the interface travels slower than the reflected wave in medium~1 but faster than the later-backward and later-forward waves in medium~2, allowing three scattered waves to coexist; we refer to this scenario as Case~I. In the comoving case, where the interface moves forward [right panel of Fig.~\ref{fig:regimes}(c)], the interface velocity exceeds the wave velocity in medium~2 and only a single scattered wave appears in medium~1; this scenario is referred to as Case~II.
In the following, we present the related frequency relations and partial solutions for the scattering coefficients, both to summarize previous attempts and to serve as a benchmark for the general solution to be introduced later for these two interluminal cases.

\subsection{Frequency Relations}

\begin{figure}[ht!]
    \centering
    \includegraphics[width=8.6cm]{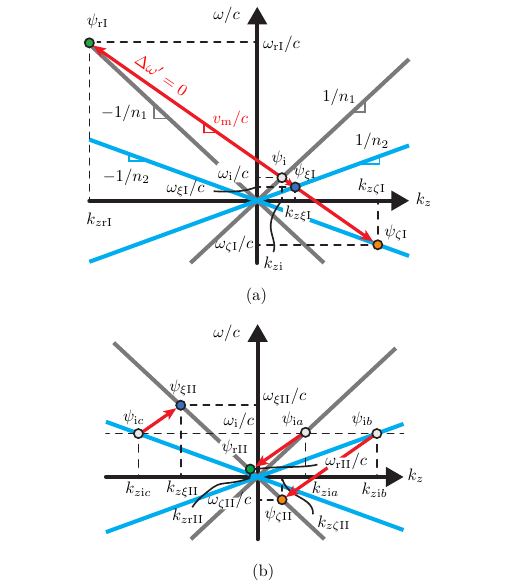}
    \caption{\label{fig:dispersion} Dispersion diagrams showing the frequency transitions at interluminal interfaces [Fig.~\ref{fig:regimes}(c)] for (a)~Case I ($v_{\m}<0$) and (b)~Case II ($v_{\m}>0$).}
\end{figure}

The frequency relations in the interluminal regime can be obtained following the same procedure as in the subluminal and superluminal regimes~\cite{Deck_2019_uniform,Caloz_2019_spacetime2}, with guidance from the dispersion diagram in Fig.~\ref{fig:dispersion}.
In the comoving frame, $K'$, which moves at the same velocity as the interface, the interface is stationary and the frequency is conserved, i.e., $\Delta\omega'=0$.
Applying the Lorentz transformation~\cite{Kong_2008_emwtheory} $\omega'=\gamma(\omega-v_{\m}k_z)$, where $\gamma=(1-v_{\m}^2/c^2)^{-1/2}$, the frequency relations in the laboratory frame, $K$, are determined as~\cite{De_2025_scattering,De_2025_dispersion}
\begin{subequations} \label{eq:w}
    \begin{equation}\label{eq:wr}
    \frac{\omega_{\mathrm{r}}}{\omega_{\mathrm{i}}}=\frac{1-v_{\m}/v_1}{1+v_{\m}/v_1},
\end{equation}
    \begin{equation}\label{eq:wb}
    \frac{\omega_{\zeta}}{\omega_{\mathrm{i}}}=\frac{1-v_{\m}/v_1}{1+v_{\m}/v_2}
\end{equation}
and
    \begin{equation}\label{eq:wtf}
    \frac{\omega_{\xi}}{\omega_{\mathrm{i}}}=\frac{1-v_{\m}/v_1}{1-v_{\m}/v_2},
\end{equation}
\end{subequations}
where $v_{1,2} = c/n_{1,2}$ are the wave velocities in the two media. These frequency relations match those known for reflected [Eq.~\eqref{eq:wr}], later-backward [Eq.~\eqref{eq:wb}] and transmitted/later-forward waves [Eq.~\eqref{eq:wtf}] in the subluminal and superluminal regimes~\cite{Deck_2019_uniform,Caloz_2019_spacetime2}.

\subsection{Scattering Coefficients for Particular Media} \label{sec:benchmarks}
\pati{Ill-Posed Problem}
The main difficulty arises when solving for the \emph{scattering coefficients} in the interluminal regime.
The moving boundary conditions~\cite{Kong_2008_emwtheory},
\begin{subequations} \label{eq:BCs}
    \begin{equation}
        \mathbf{E}_{1}+\mathbf{v}_{\m}\times\mathbf{B}_{1}=\mathbf{E}_{2}+\mathbf{v}_{\m}\times\mathbf{B}_{2}
    \end{equation}
    and 
    \begin{equation}
        \mathbf{H}_{1}-\mathbf{v}_{\m}\times\mathbf{D}_{1}=\mathbf{H}_{2}-\mathbf{v}_{\m}\times\mathbf{D}_{2}
    \end{equation}
\end{subequations}
provide the two conditions that determine the scattering coefficients.
As shown in Fig.~\ref{fig:regimes}(c), in Case~I there are then three unknowns for two conditions, leading to an underdetermined problem. In contrast, in Case~II there is one unknown for two conditions, resulting in an overdetermined problem. As a result, it seems impossible to obtain exact scattering coefficients in this regime based on the two equations given in Eqs.~\eqref{eq:BCs}.

\pati{Previous Resolution Attempts}
To address the two ill-posed interluminal problems, it was suggested to insert an intermediate graded-index slab between the two media~\cite{Ostrovskii_1967_correct,Ostrovskii_1975_some,Deck_2019_scattering}, with refractive index varying from subluminal $n_{\m}-\Delta n$ to superluminal $n_{\m}+\Delta n$, where $n_{\m}=c/|v_{\m}|=\sqrt{\epsilon_{\mathrm{r}\m}\mu_{\mathrm{r}\m}}$ and $\Delta n\to0$. In this approach, the ill-determined interluminal problem is decomposed into two well-determined problems, a subluminal one for $n\in(n_{\m}-\Delta n,n_{\m})$ and a superluminal one for $n\in(n_{\m},n_{\m}+\Delta n)$, and the slab is ultimately reduced to be infinitesimally thin to recover the result for an actual interluminal interface. This leads to the following results for Cases~I and~II~\cite{Deck_2019_scattering}:

\begin{equation} \label{eq:eta_m}
\begin{gathered}
    \mathrm{r}_{\mathrm{I}}=\frac{\eta_{\m}-\eta_1}{\eta_1+\eta_{\m}}\frac{1-v_{\m}/v_1}{1+v_{\m}/v_1},\quad \mathrm{r}_{\mathrm{II}}=\frac{\eta_{\m}-\eta_1}{\eta_1+\eta_{\m}}\frac{1-v_{\m}/v_1}{1+v_{\m}/v_1}, \\
    \zeta_{\mathrm{I}}=\frac{\eta_{\m}-\eta_2}{\eta_1+\eta_{\m}}\frac{1-v_{\m}/v_1}{1+v_{\m}/v_2},\quad \zeta_{\mathrm{II}}=\frac{\eta_1}{\eta_2}\frac{\eta_{2}-\eta_{\m}}{\eta_1+\eta_{\m}}\frac{1-v_{\m}/v_2}{1+v_{\m}/v_1}, \\
    \xi_{\mathrm{I}}=\frac{\eta_{\m}+\eta_2}{\eta_1+\eta_{\m}}\frac{1-v_{\m}/v_1}{1-v_{\m}/v_2},\quad \xi_{\mathrm{II}}=\frac{\eta_1}{\eta_2}\frac{\eta_{2}+\eta_{\m}}{\eta_1+\eta_{\m}}\frac{1+v_{\m}/v_2}{1+v_{\m}/v_1}.
\end{gathered}
\end{equation}

Compared to the actual interluminal parameters [Fig.~\ref{fig:regimes}(c)], the scattering coefficients in Eqs.~\eqref{eq:eta_m} involve an additional parameter, the wave impedance of the intermediate slab $\eta_{\m}=\eta_0 \sqrt{\mu_{\mathrm{r}\m}/\epsilon_{\mathrm{r}\m}}$, where $\eta_0$ is the impedance of free space. For the technique to provide an acceptable result, this parameter must vanish upon taking the zero-thickness limit. Unfortunately, that is not the case in general.
Previous studies considered the particular case of \emph{non-magnetic media}, where $\mu_{\mathrm{r}1}=\mu_{\mathrm{r}2}=\mu_{\mathrm{rm}}=1$, allowing the unknown impedance to be expressed in terms of $v_{\m}$ as $\eta_{\m}=\eta_0/n_{\m}=|v_{\m}|\eta_0/c$, and hence effectively disappearing from the equation to yield the exact solution for this scenario.
In fact, Eqs.~\eqref{eq:eta_m} also apply to two non-considered additional cases: \emph{non-electric media}, where $\epsilon_{\mathrm{r}1}=\epsilon_{\mathrm{r}2}=\epsilon_{\mathrm{rm}}=1$ (yielding $\eta_{\m}=c\eta_0/|v_{\m}|$) and \emph{impedance-matched media}, where $\mu/\epsilon=\mathrm{const.}$ (yielding $\eta_{\m}=\eta_1=\eta_2$). Table~\ref{tab:benchmarks} summarizes the solutions for these three cases.
However, in the general case where $\mu_{\mathrm{r}\m}$ and $\epsilon_{\mathrm{r}\m}$ vary independently, the impedance $\eta_{\m}$ is not uniquely defined and Eqs.~\eqref{eq:eta_m} cannot be applied.

\begin{table*}[t]
\centering
\caption{Interluminal scattering coefficients for particular cases.}
\begin{tabular}{Sc|Sccc|Sccc}
\hline
\multirow{2}{*}{\textbf{Configuration}} 
& \multicolumn{3}{c|}{\begin{minipage}{6cm}\centering \textbf{Case~I} ($v_{\m}<0$)\end{minipage}} 
& \multicolumn{3}{c}{\begin{minipage}{6cm}\centering \textbf{Case~II} ($v_{\m}>0$)\end{minipage}} \\
\cline{2-7}
& \begin{minipage}{2cm}\centering $\mathrm{r}_{\mathrm{I}}$ \end{minipage} & \begin{minipage}{2cm}\centering $\zeta_{\mathrm{I}}$ \end{minipage} & \begin{minipage}{2cm}\centering $\xi_{\mathrm{I}}$ \end{minipage} 
& \begin{minipage}{2cm}\centering $\mathrm{r}_{\mathrm{II}}$ \end{minipage} & \begin{minipage}{2cm}\centering $\zeta_{\mathrm{II}}$ \end{minipage} & \begin{minipage}{2cm}\centering $\xi_{\mathrm{II}}$ \end{minipage} \\
\hline
\
\shortstack{Non-magnetic \\ ($\mu_{\mathrm{r}1}=\mu_{\mathrm{r}2}=1$)} 
& $-1$ & $-\frac{v_2}{v_1}$ & $\frac{v_2}{v_1}$
& $-\left(\frac{1-v_{\m}/v_1}{1+v_{\m}/v_1}\right)^2$ & $\left(\frac{1-v_{\m}/v_2}{1+v_{\m}/v_1}\right)^2$ & $\left(\frac{1+v_{\m}/v_2}{1+v_{\m}/v_1}\right)^2$ \\

\shortstack{Non-electric \\ ($\epsilon_{\mathrm{r}1}=\epsilon_{\mathrm{r}2}=1$)}
& $1$ & $1$ & $1$
& $\left(\frac{1-v_{\m}/v_1}{1+v_{\m}/v_1}\right)^2$ & $-\frac{v_2}{v_1}\left(\frac{1-v_{\m}/v_2}{1+v_{\m}/v_1}\right)^2$ & $\frac{v_2}{v_1}\left(\frac{1+v_{\m}/v_2}{1+v_{\m}/v_1}\right)^2$ \\

\shortstack{Impedance-matched \\ ($\mu/\epsilon=\mathrm{const.}$)}
& $0$ & $0$ & $\frac{1-v_{\m}/v_1}{1-v_{\m}/v_2}$
& $0$ & $0$ & $\frac{1+v_{\m}/v_1}{1+v_{\m}/v_2}$ \\
\hline
\end{tabular}
\label{tab:benchmarks}
\end{table*}

\section{General Solution} \label{sec:method}

In this section, we present a symmetric decomposition approach with detailed derivations for the general interluminal scattering coefficients, using Case~I as an example; Case~II can be derived similarly.

\subsection{Symmetric Decomposition}

Decomposing a continuous interface into a series of discrete sub-interfaces and taking the limit as each step becomes infinitesimally small provides an effective approach to solve the scattering problem at such an interface~\footnote{The relevant quantity here is the \emph{local slope} of the interface in the space-time diagram rather than that of the total length. The total geometric length of a continuous curve cannot, in general, be exactly recovered through polygonal discretization---an issue analogous to the classical staircase paradox~\cite{Moscovich_2006_Loopy}}.

\begin{figure}[ht!]
    \centering
    \includegraphics[width=8.6cm]{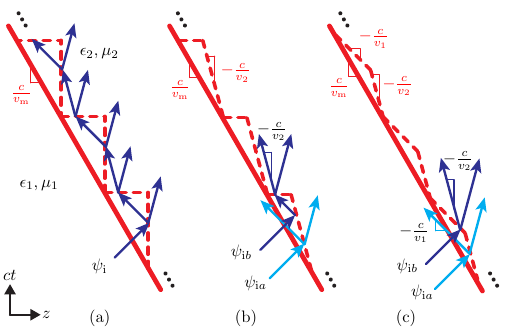}
    \caption{\label{fig:polylines} Different polyline decompositions of a contramoving ($v_{\m}<0$) interluminal alternating and corresponding wave scattering. (a)~Staircase decomposition combining pure-space and pure-time interfaces. (b)~Asymmetric subluminal-limit and pure-time decomposition~\cite{Shui_2014_one}. (c)~Symmetric decomposition combining subluminal-limit and superluminal-limit interfaces. The red solid line indicates the target interface trajectory and the red dashed lines represent its polyline approximations.}
\end{figure}

Several decomposition approaches are illustrated in Fig.~\ref{fig:polylines}.
The simplest approach is the staircase decomposition [Fig.~\ref{fig:polylines}(a)]~\cite{Wu_2025_electromagnetic}, consisting of a pure-space interface and a pure-time interface. This method is commonly used in numerical techniques such as the finite-difference time-domain (FDTD) method~\cite{Taflove_2005_FDTD,Bahrami_2023_FDTD}. Although straightforward, it unfortunately introduces spurious---unphysical---multiple scattering at each sub-interface, as shown by arrows in the figure.
Another approach combines a subluminal-limit interface with velocity $-v_2$ and a pure-time interface [Fig.~\ref{fig:polylines}(b)]. This method has been applied to the acoustic ``intersonic'' related problem~\cite{Shui_2014_one}. Compared with the staircase decomposition [Fig.~\ref{fig:polylines}(a)], this approach eliminates interference between waves incident on different sub-interfaces. However, it still support spurious multiple scattering at the pure-time interface [dark-blue trajectories in Fig.~\ref{fig:polylines}(b)], obscuring the underlying physics.
To properly model the physics of the problem, we resort to the only decomposition that avoids spurious multiple scattering. That is the symmetric decomposition shown in Fig.~\ref{fig:polylines}(c), combines a subluminal-limit interface, with velocity $-v_2$, and a superluminal-limit interface, with velocity $-v_1$. In this configuration, each wave interacts with the interfaces only once, as is the case in reality.

\subsection{Infinitesimal Limit}
\pati{Different Regions}
The proposed symmetric decomposition method [Fig.~\ref{fig:polylines}(c)] transforms the previously unresolved interluminal scattering problem into a sequence of sound and well-established subluminal and superluminal cases. To recover the exact scattering coefficients of a continuous interluminal interface, after computing the separate scattering parameters, we reduce the size of the discretized step to the infinitesimal limit, so as to find the sough after scattering coefficients.

\begin{figure}[ht!]
    \centering
    \includegraphics[width=8.6cm]{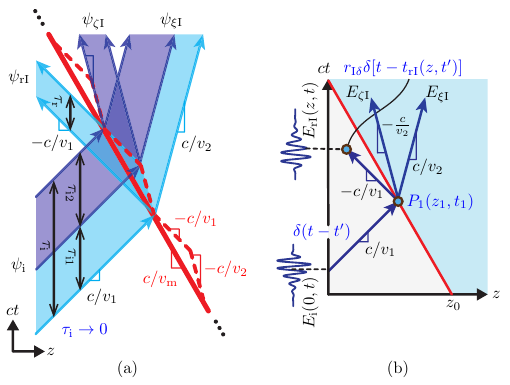}
    \caption{\label{fig:impulse} General solution for a contramoving interluminal interface ($v_{\m}<0$) in Fig.~\ref{fig:polylines}(c). (a)~Symmetric decomposition within a single step, with $\tau_{\mathrm{i}}\to 0$. The light-blue region indicates the portion of the wave interacting with the subluminal-limit interface, while the dark-blue region indicates the portion interacting with the superluminal-limit interface. (b)~Extension to arbitrary incidence using the space-time impulse response method, with $z_{0}$ denoting the initial position of the interface at $t=0$.}
\end{figure}

Figure~\ref{fig:impulse}(a) illustrates the scattering process within a single step of the symmetrical decomposition shown in Fig.~\ref{fig:polylines}(c).
An incident square pulse wave with infinitesimal duration $\tau_{\mathrm{i}}\to~0$ interacting with this step is divided into two regions, distinguished by colors: the light-blue region corresponds to the portion of the wave interacting with the subluminal-limit interface ($-v_2$) and the dark-blue region corresponds to the portion interacting with the superluminal-limit interface ($-v_1$).
As shown in Fig.~\ref{fig:impulse}(a), the wave originating from the light-blue region splits into a reflected wave in medium~1 and a transmitted wave in medium~2, with the (subluminal) scattering coefficients
\begin{equation} \label{eq:r_t_0}
r_{12}=\frac{\eta_2-\eta_1}{\eta_1+\eta_2} \quad\text{and}\quad
t_{12}=\frac{2\eta_2}{\eta_1+\eta_2},
\end{equation}
while the wave originating from the dark-blue region splits into later-backward and later-forward waves in medium~2, with the (superluminal) scattering coefficients
\begin{equation} \label{eq:zeta_xi_0}
\zeta_{12}=\frac{\eta_1-\eta_2}{2\eta_1}\quad\text{and}\quad
\xi_{12}=\frac{\eta_1+\eta_2}{2\eta_1}.
\end{equation}
Note that Eqs.~\eqref{eq:r_t_0} and~\eqref{eq:zeta_xi_0} correspond to the scattering coefficients of a \emph{stationary} interface.
This is because the incident pulse considered here is infinitesimally short and therefore does not experience any motion of the interface during the scattering interaction~\footnote{A similar principle of impulsive scattering in the limiting case of pure-time modulation, with $v_{\m}\to\pm\infty$, can be found in~\cite{Xiao_2014_reflection}.}.

\pati{Weak Solution or Weighted Averaging}
In the limit $\tau_{\mathrm{i}} \to 0$, the wave amplitude can be considered constant during the interaction. A weak-solution---or weighted-averaging---method~\cite{Shui_2014_one} may then be applied to derive the effective scattering coefficients for such an impulsive incidence at an interluminal interface.
The first step is to determine the weighting parameter, representing the fraction of the wave corresponding to the different colored regions.
The fraction of the light-blue incidence, denoted $\tau_{\mathrm{i}1}$, can be obtained from duration relations~\cite{Caloz_2019_spacetime2,Deck_2019_uniform}:
\begin{equation}
\frac{\tau_{\mathrm{r}}}{\tau_{\mathrm{i}}}=\frac{1+v_{\m}/v_1}{1-v_{\m}/v_1}
\quad \text{and} \quad
\frac{\tau_{\mathrm{r}}}{\tau_{\mathrm{i}1}}=\frac{1-v_2/v_1}{1+v_2/v_1},
\end{equation}
which yields the weighting ratio
\begin{equation} \label{eq:a1}
\alpha_1=\frac{\tau_{\mathrm{i}1}}{\tau_{\mathrm{i}}}
=\frac{1+v_{\m}/v_1}{1-v_{\m}/v_1}
\frac{1+v_2/v_1}{1-v_2/v_1}.
\end{equation}
The effective scattering coefficients for the impulsive incidence in Case~I are then obtained by using weighted averaging as
\begin{equation} \label{eq:scattering_delta}
\resizebox{\linewidth}{!}{$
\begin{gathered}
r_{\mathrm{I}\delta}=\lim_{\tau_{\mathrm{i}}\to 0} \frac{r_{12}\int_{\tau_{\mathrm{i}1}} \psi_{\mathrm{i}} ~\mathrm{d}t}{\int_{\tau_{\mathrm{i}}} \psi_{\mathrm{i}}~\mathrm{d}t}=\alpha_1 r_{12},\\
\zeta_{\mathrm{I}\delta}=\lim_{\tau_{\mathrm{i}}\to 0} \frac{\zeta_{12}\int_{\tau_{\mathrm{i}2}} \psi_{\mathrm{i}}~\mathrm{d}t}{\int_{\tau_{\mathrm{i}}} \psi_{\mathrm{i}}~\mathrm{d}t}=(1-\alpha_1)\zeta_{12},\\
t_{\mathrm{I}\delta}=\lim_{\tau_{\mathrm{i}}\to 0} \frac{t_{12}\int_{\tau_{\mathrm{i}1}} \psi_{\mathrm{i}}~\mathrm{d}t+\xi_{12}\int_{\tau_{\mathrm{i}2}} \psi_{\mathrm{i}}~\mathrm{d}t}{\int_{\tau_{\mathrm{i}}} \psi_{\mathrm{i}}~\mathrm{d}t}=\alpha_1 t_{12}+(1-\alpha_1)\xi_{12},
\end{gathered}$}
\end{equation}
where $\alpha_1$ is given in Eq.~\eqref{eq:a1}, and $r_{12}$, $t_{12}$, $\zeta_{12}$ and $\xi_{12}$ are provided in Eqs.~\eqref{eq:r_t_0} and~\eqref{eq:zeta_xi_0}.
The subscript $\delta$ denotes the coefficients corresponding to the impulsive incidence.

\subsection{Physicality of the Proposed Scheme}
\begin{figure}[ht!]
    \centering
    \includegraphics[width=8.6cm]{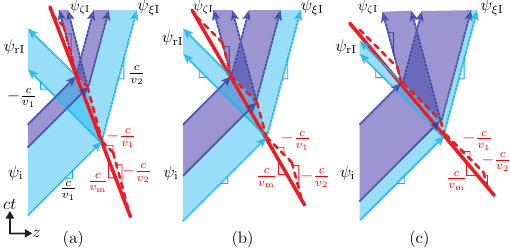} \caption{\label{fig:diff_vms} Application of the $-v_2$-subluminal and $-v_1$-superluminal decomposition of Fig.~\ref{fig:impulse}(a) to different modulation velocities, with $v_1=c$ and $v_2=0.27c$. (a)~$v_{\m}=-0.38c$. (b)~$v_{\m}=-0.58c$. (c)~$v_{\m}=-0.84c$.}
\end{figure}

Our approach of interluminality, combining the symmetric decomposition in Fig.~\ref{fig:polylines}(c) and the weighted averaging in Fig.~\ref{fig:impulse}(a), seems to resolve the enigma of the related physics. As noted in Sec.~\ref{sec:benchmarks}, the interluminal regime is ill-determined---under-determined in Case~I and over-determined in Case~II. Such indetermination is of \emph{macroscopic nature}, since it results from the application of the macroscopic moving boundary conditions of Eqs.~\eqref{eq:BCs}. This suggests the existence of a \emph{hidden, microscopic mechanism} that ``tells'' the incident wave how to deal with the extra (Case~I) or missing (Case~II) piece of information. According to our approach, interluminality is in fact a \emph{hybrid subluminal--superluminal phenomenon} whereby the incident wave is automatically decomposed by the microscopic nature of the interface into a subluminal channel and a superluminal channel, just as in the case of circular birefringence, for instance, where an incident linearly polarized wave is automatically split by the microscopic nature of the medium into right circularly polarized (RCP) and left
circularly polarized (LCP) channels. In this sense, the subluminal and superluminal channels would form the \emph{eigenstates} of interluminality. Note that this ``eigenstate decomposition'' perfectly accounts for different modulation velocities ($v_\textrm{m}$) which, as illustrated in Fig.~\ref{fig:diff_vms}, would correspond to different splitting ratios of the incident wave into subluminal and superluminal scattering. This explanation clearly pertains to Case~I, but it also applies to Case~II. Indeed, upon adding two incident waves as shown in the right panel of Fig.~\ref{fig:regimes}(c), Case~II is nothing but a time-reversed version of Case~I (to be further discussed in Sec.~\ref{sec:time_rev}), where the separate forward and backward incident waves recombine into a unique backward wave after the interface, as opposite CP waves recombine into a linearly polarized wave at the output of a circularly birefringent crystal.

\subsection{Extension to Arbitrary Waveforms}

In this section, we extend the impulsive solution given in Eq.~\eqref{eq:scattering_delta} to arbitrary incident field $E(0,t)$ waveforms using the space-time impulse response method developed in~\cite{Xiao_2011_spectral,Li_2025_Chirp} [Fig.~\ref{fig:impulse}(b)], according to which
\begin{equation} \label{eq:convolution}
        E(z,t)=\int^{\infty}_{-\infty} h(z,t,t')E(0,t')~\mathrm{d}t',
\end{equation}
where $E(z,t)$ is the output field and $h(z,t,t')$ is the corresponding impulse response.

As shown in Fig.~\ref{fig:impulse}(b), an impulse launched at time $t'$ propagates with velocity $v_1$ in medium~1 and interacts with the interface at the point $P_1(z_1,t_1)$. The coordinates of this point are determined from the space-time geometry as
\begin{equation} \label{eq:P1}
    z_1=\frac{v_{\m}t'+z_0}{1-v_{\m}/v_1}\quad \text{and} \quad t_1=\frac{t'+z_0/v_1}{1-v_{\m}/v_1}.
\end{equation}
Upon interacting with the interface, the impulse splits into three components. For the reflected wave, the impulse undergoes an amplitude change given by the scattering coefficient $r_{\mathrm{I}\delta}$ [Eq.~\eqref{eq:scattering_delta}] and a reversal of velocity to $-v_1$. At position $z$, the arrival time of the reflected wave $t_{\mathrm{rI}}$ satisfies the space-time relation
\begin{equation} \label{eq:tr1}
    z-z_1=-v_1(t_{\mathrm{rI}}-t_1).
\end{equation}
Hence, the reflection impulse response is
\begin{subequations} \label{eq:hr1}
    \begin{equation} 
h_{\mathrm{rI}}(z,t,t') = r_{\mathrm{I}\delta}\delta[t - t_{\mathrm{rI}}(z,t')],
\end{equation}
where
\begin{equation}
    t_{\mathrm{rI}}(z,t')=\frac{1+v_{\m}/v_1}{1-v_{\m}/v_1}t'-\frac{z}{v_1}+\frac{2z_0/v_1}{1-v_{\m}/v_1}
\end{equation}
\end{subequations}
is obtained upon substituting Eq.~\eqref{eq:P1} into Eq.~\eqref{eq:tr1} and solving for $ t_{\mathrm{rI}}$.
Finally, substituting Eq.~\eqref{eq:hr1} into Eq.~\eqref{eq:convolution} and evaluating the resulting integral, the reflected field is obtained as
\begin{subequations} \label{eq:E_rI}
    \begin{equation}
E_{\mathrm{rI}}(z,t)=r_{\mathrm{I}}E\left[0,\frac{1-v_{\m}/v_1}{1+v_{\m}/v_1}\left(t+\frac{z}{v_1}\right)-\frac{2 z_0/v_1}{1+v_{\m}/v_1}\right],
\end{equation}
where
\begin{equation}
r_{\mathrm{I}}=r_{\mathrm{I}\delta}\frac{1-v_{\m}/v_1}{1+v_{\m}/v_1}=\frac{\eta_2-\eta_1}{\eta_1+\eta_2}\frac{1+v_2/v_1}{1-v_2/v_1}
\end{equation}
is the generalized reflected coefficient.
\end{subequations}
Similarly, the later-backward field is found as
\begin{subequations} \label{eq:E_zetaI}
    \begin{equation}
    \resizebox{\linewidth}{!}{$
E_{\zeta\mathrm{I}}(z,t)=\zeta_{\mathrm{I}}E\left[0,\frac{1-v_{\m}/v_1}{1+v_{\m}/v_2}\left(t+\frac{z}{v_2}\right)-\frac{(1/v_1+1/v_2) z_0}{1+v_{\m}/v_2}\right],$}
\end{equation}
where
\begin{equation}
\zeta_{\mathrm{I}}=\frac{\eta_2-\eta_1}{\eta_1}\frac{v_2}{v_1-v_2}
\end{equation}
is the generalized later-backward-wave coefficient,
\end{subequations}
and the later-forward field as
\begin{subequations} \label{eq:E_xiI}
    \begin{equation}
    \resizebox{\linewidth}{!}{$
E_{\xi\mathrm{I}}(z,t)=\xi_{\mathrm{I}}E\left[0,\frac{1-v_{\m}/v_1}{1-v_{\m}/v_2}\left(t-\frac{z}{v_2}\right)-\frac{(1/v_1-1/v_2) z_0}{1-v_{\m}/v_2}\right],$}
\end{equation}
where
\begin{equation}
\xi_{\mathrm{I}}=\frac{(\eta_1^2+\eta_2^2)(1+v_{\m}/v_2)-2\eta_1\eta_2(v_1/v_2+v_{\m}/v_1)}{\eta_1(\eta_1+\eta_2)(1-v_1/v_2)(1-v_{\m}/v_2)}
\end{equation}
is the generalized later-forward-wave coefficient.
\end{subequations}

Same derivations apply to Case~II, yielding
\begin{subequations} \label{eq:E_II}
\begin{equation}
E_{\mathrm{rII}}(z,t) = r_{\mathrm{II}}E\left[0,\frac{1-v_{\m}/v_1}{1+v_{\m}/v_1}\left(t+\frac{z}{v_1}\right)-\frac{2 z_0/v_1}{1+v_{\m}/v_1}\right],
\end{equation}
\begin{equation}
\begin{aligned}
E_{\zeta\mathrm{II}}(z,t)=\zeta_{\mathrm{II}}E\Bigg[0,-\frac{1-v_{\m}/v_2}{1+v_{\m}/v_1}&\Bigg(t+\frac{z}{v_1}\Bigg)\\+&\frac{(1/v_1+1/v_2) z_0}{1+v_{\m}/v_1}\Bigg],
\end{aligned}
\end{equation}
and
\begin{equation}
\begin{aligned}
E_{\xi\mathrm{II}}(z,t)=\xi_{\mathrm{II}}E\Bigg[0,-\frac{1+v_{\m}/v_2}{1+v_{\m}/v_1}&\Bigg(t-\frac{z}{v_1}\Bigg)\\+&\frac{(1/v_2-1/v_1) z_0}{1+v_{\m}/v_1}\Bigg],
\end{aligned}
\end{equation}
with the corresponding generalized scattering coefficients
\begin{equation}
r_{\mathrm{II}}=\frac{\eta_2-\eta_1}{\eta_1+\eta_2}\frac{1+v_2/v_1}{1-v_2/v_1}\left(\frac{1-v_{\m}/v_1}{1+v_{\m}/v_1}\right)^2,
\end{equation}
\begin{equation}
\zeta_{\mathrm{II}}=\frac{\eta_2-\eta_1}{\eta_2}\frac{v_2}{v_2-v_1}\left(\frac{1-v_{\m}/v_2}{1+v_{\m}/v_1}\right)^2,
\end{equation}
and
\begin{equation}
\resizebox{\linewidth}{!}{$
\xi_{\mathrm{II}}=\frac{(1+v_{\m}/v_2)[(\eta_1^2+\eta_2^2)(1-v_{\m}/v_2)-2\eta_1\eta_2(v_1/v_2-v_{\m}/v_1)]}{\eta_2(\eta_1+\eta_2)(1-v_1/v_2)(1+v_{\m}/v_1)^2}.$}
\end{equation}
\end{subequations}

\section{Validation}

\pati{Particular Media}
The general solutions in Eqs.~\eqref{eq:E_rI}-\eqref{eq:E_II} can be benchmarked against the particular cases listed in Table~\ref{tab:benchmarks}. It may be easily verified that substituting the corresponding impedance relations---$\eta_{1,2}=\eta_0/n_{1,2}$ for the non-magnetic case, $\eta_{1,2}=\eta_0 n_{1,2}$ for the non-electric case and $\eta_{1}/\eta_2=1$ for the impedance-matched case---into the scattering coefficients indeed reproduces these results.

\begin{figure}[ht!]
    \centering
    \includegraphics[width=8.6cm]{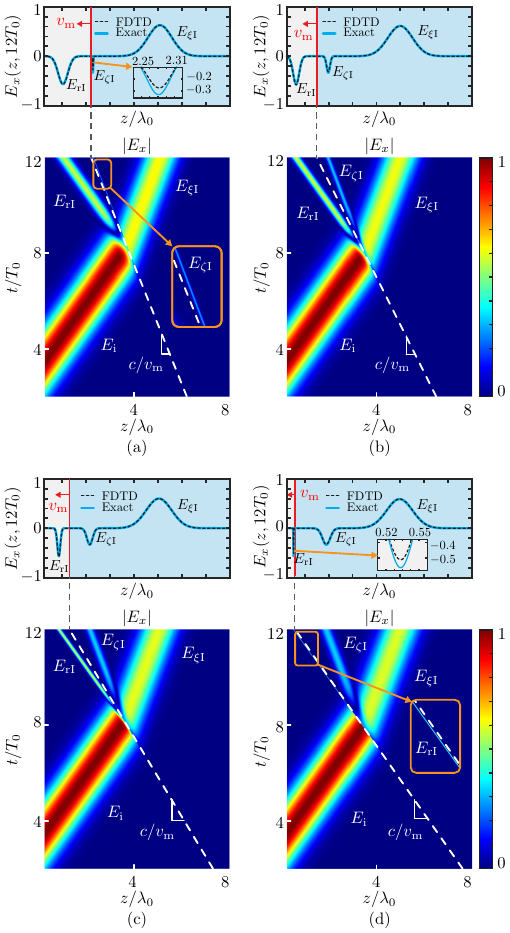}
    \caption{\label{fig:FDTD} Full-wave (FDTD) simulations of a Gaussian pulse, $E_{x\mathrm{i}}(0,t)=\mathrm{e}^{-(t-3.5T_0)^2/2T_0^2}$, scattering at a contramoving interface with $\epsilon_{\mathrm{r}1}=1.3$, $\mu_{\mathrm{r}1}=1.5$, $\epsilon_{\mathrm{r}2}=3.5$ and $\mu_{\mathrm{r}2}=2$ in the interluminal regime ($-0.38c<v_{\m}<-0.72c$). The modulation velocities are (a) $v_{\m}=-0.4c$, (b) $v_{\m}=-0.5c$, (c) $v_{\m}=-0.6c$ and (d) $v_{\m}=-0.7c$. The bottom panels show the electric-field magnitude $|E_x|$ in space-time diagrams normalized to the free-space period $T_0$ and wavelength $\lambda_0 = cT_0$, where the white dashed lines indicate the interface trajectories. The top panels show the corresponding fields at $t=12T_0$.}
\end{figure}

\pati{FDTD}
In addition to this analytical benchmarking, FDTD simulations~\cite{Deck_2023_yeecell,Bahrami_2023_FDTD} provide an independent validation approach.
Figure~\ref{fig:FDTD} shows a Gaussian pulse scattering at a contramoving modulation interface (Case I) for different interluminal velocities (See Appendix~\ref{sec:Case_II} for Case II). The top panels compare the analytical results [Eqs.~\eqref{eq:E_rI}-\eqref{eq:E_xiI}] with the FDTD fields at $t=12T_0$, showing good agreement.
Larger numerical values in the peak amplitudes for the later-backward wave in Fig.~\ref{fig:FDTD}(a) and reflected wave in Fig.~\ref{fig:FDTD}(d) (insets) are due to insufficient spatial resolution and numerical dispersion in FDTD for the extreme compression, occurring close to the corresponding interluminal limits.

\begin{figure}[ht!]
    \centering
    \includegraphics[width=8.6cm]{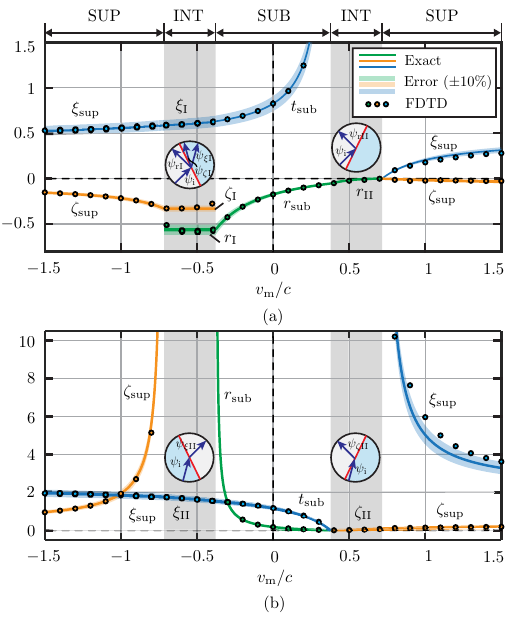}
    \caption{\label{fig:limits} Scattering coefficients versus the normalized modulation velocity $v_{\m}/c$ for the incidence from (a)~medium 1 and (b)~medium 2. The media parameters are identical to those used in Fig.~\ref{fig:FDTD}. `SUB', `INT' and `SUP' at the top of the graph denote the subluminal, interluminal and superluminal regimes, respectively. $r_{\mathrm{sub}}$ and $t_{\mathrm{sub}}$ are the subluminal reflection and transmission coefficients; $\zeta_{\mathrm{sub}}$ and $\xi_{\mathrm{sub}}$ are the superluminal later-backward and later-forward coefficients, as given in~\cite{Caloz_2019_spacetime2,Deck_2019_uniform}.}
\end{figure}

\pati{Scattering Coefficients vs Velocity}
Figure~\ref{fig:limits} shows the scattering coefficients as a function of the normalized modulation velocity $v_{\m}/c$ for the same media parameters as in Fig.~\ref{fig:FDTD}, covering the subluminal, interluminal and superluminal regimes, with FDTD results providing validation. For clarity, the results are divided into two cases: Fig.~\ref{fig:limits}(a) corresponds to incidence from the rarer medium (medium 1), while Fig.~\ref{fig:limits}(b) corresponds to incidence from the denser medium (medium 2). As shown in Fig.~\ref{fig:limits}(a), the interluminal scattering coefficients for Case I connect smoothly with the subluminal and superluminal limits, similar to Case II in the comoving regime shown in Fig.~\ref{fig:limits}(a) and Fig.~\ref{fig:limits}(b). This continuity provides a further validation of the theory.

%

\section{Further Interesting Physics}
\subsection{Velocity-Independent Coefficients in Case I}

An interesting observation in the left part of Fig.~\ref{fig:limits}(a) is that, in Case~I, the interluminal reflection and later-backward coefficients are constant, i.e., independent of $v_{\m}$, and exhibit the same values as those at the subluminal and superluminal limits, respectively: $r_{\mathrm{I}}=r_{\mathrm{sub}}(v_{\m}=-v_2)$ and $\zeta_{\mathrm{I}}=\zeta_{\mathrm{sup}}(v_{\m}=-v_1)$, where $r_{\mathrm{sub}}=\frac{\eta_2-\eta_1}{\eta_1+\eta_2}\frac{1-v_{\m}/v_1}{1+v_{\m}/v_1}$ denotes the subluminal reflection coefficient and $\zeta_{\mathrm{sup}}=\frac{\eta_2-\eta_1}{2\eta_1}\frac{1-v_{\m}/v_1}{1+v_{\m}/v_2}$ denotes the superluminal later-backward coefficient~\cite{Caloz_2019_spacetime2,Deck_2019_uniform}.
This behavior can be understood from Fig.~\ref{fig:diff_vms}. Within the symmetric decomposition, interluminal reflection always originates from the subluminal-limit ($-v_2$) interface, while later-backward scattering always originates from the superluminal-limit ($-v_1$) interface, independent of the actual interluminal velocity $v_{\m}$. In contrast, the later-forward wave arises from a combination of both contributions, explaining why the former two coefficients are velocity-independent, whereas the latter depends on~$v_{\m}$.

\subsection{Shock Wave in Case II}\label{sec:shock_waves}

\begin{figure}[ht!]
    \centering
    \includegraphics[width=8.6cm]{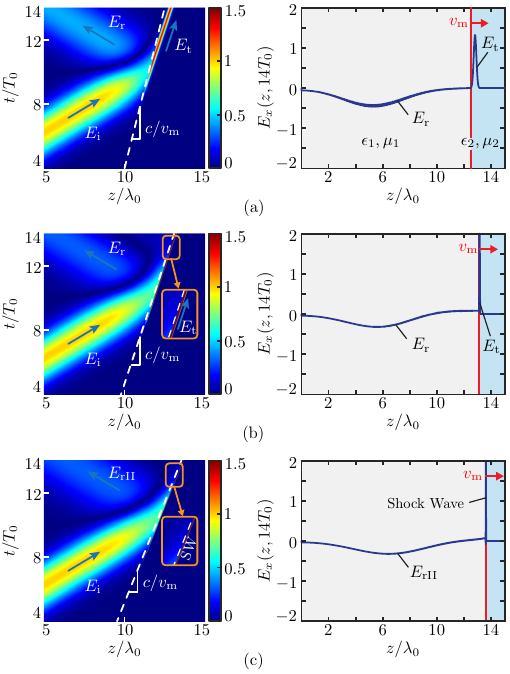}
    \caption{\label{fig:shock} FDTD illustration of shock-wave formation in Case II. The media parameters are $\epsilon_{\mathrm{r}1}=0.3$, $\mu_{\mathrm{r}1}=1.2$, $\epsilon_{\mathrm{r}2}=6$, and $\mu_{\mathrm{r}2}=1.5$. The interluminal regime corresponds to $0.333c<v_{\m}<1.667c$. The interface velocities are set to (a) a subluminal value $v_{\m}=0.25c$, (b) the subluminal-limit value $v_{\m}=0.33c$ and (c) an interluminal value $v_{\m}=0.4c$, respectively. The left panels show the electric-field magnitude $|E_x|$ in space-time diagrams, normalized to $T_0$ and $\lambda_0$, with a Gaussian incident pulse $E_{x\mathrm{i}}(0,t) = \mathrm{e}^{-(t-3.5T_0)^2 / 2T_0^2}$. Dashed white lines indicate the interface trajectories and `SW' denotes the shock wave. The right panels show the corresponding fields at $t = 14T_0$.}
\end{figure}

Another interesting note is that the scattering coefficients of Case I satisfy the moving boundary conditions [Eq.~\eqref{eq:BCs}] \emph{and} the energy-momentum relations in~\cite{Li_2025_Energy} (see Appendix~\ref{sec:energy}), whereas those of Case~II do not. This deviation from the conventional scattering picture in Case~II arises because the extreme motion of the interface generates not only a regular scattered wave but also a \emph{shock wave} localized at the interface~\cite{Ostrovskii_1975_some,Shui_2014_one} and corresponding to a singularity that is not accounted for in~\cite{Li_2025_Energy}.
Figure~\ref{fig:shock} illustrates such shock-wave formation in Case~II as the modulation velocity increases from subluminal to interluminal. As shown in Figs.~\ref{fig:shock}(a) and (b), when the interface velocity rises from a subluminal value to the subluminal-limit velocity, the transmitted wave is strongly compressed due to the ``pushing effect'' of the interface, with the local wavelength approaching zero. Increasing the interface velocity further into the interluminal regime [Fig.~\ref{fig:shock}(c)] induces the formation of a shock wave characterized by an extremely short local wavelength. This extreme compression at the interface results in field accumulation and wavefront piling-up, resembling the physics of a supersonic boom.

\subsection{Time-Reversal Symmetry}\label{sec:time_rev}

\begin{figure}[ht!]
    \centering
    \includegraphics[width=8.6cm]{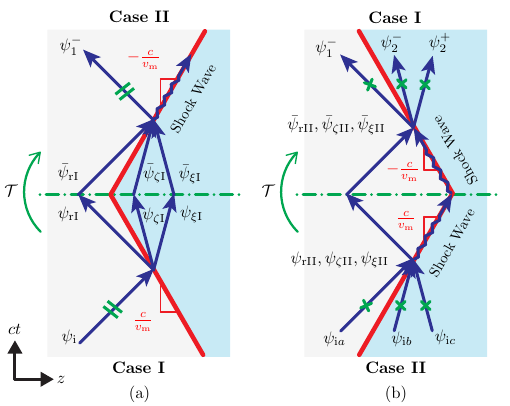}
    \caption{\label{fig:TR} Time-reversal symmetry for (a)~Case I and time-reversal symmetry breaking for (b)~Case II.}
\end{figure}

The existence of the shock wave in Case II also explains the time-reversal paradox observed in Fig.~\ref{fig:TR}.
As shown in Fig.~\ref{fig:TR}(a), the scattering coefficients satisfy the time-reversal relation
\begin{equation} \label{eq:TR1}
r_{\mathrm{I}}\bar{r}_{\mathrm{II}} + \zeta_{\mathrm{I}}\bar{\zeta}_{\mathrm{II}} + \xi_{\mathrm{I}}\bar{\xi}_{\mathrm{II}} = 1,
\end{equation}
where the overbarred quantities denote time-reversed quantities, i.e, $\bar{\psi}=\psi(-v_{\m})$.
In contrast, for Fig.~\ref{fig:TR}(b),
\begin{equation} \label{eq:TR2}
(r_{\mathrm{II}} + \zeta_{\mathrm{II}} + \xi_{\mathrm{II}}) \bar{r}_{\mathrm{I}} \neq 1.
\end{equation}
The violation of time-reversal symmetry in Eq.~\eqref{eq:TR2} arises from the time-reversal component of the shock-wave singularity, which interferes with the reflected wave in Case I [Fig.~\ref{fig:TR}(b)] but does not affect it in Case II [Fig.~\ref{fig:TR}(a)].

%

\section{Conclusion}

In this work, we have resolved the interluminal scattering problem and established a general formulation for this regime. Our analysis shows that the three scattered waves in the contramoving interluminal case can be traced back to the behavior of the subluminal and superluminal limit interfaces. We also find that a shock wave appears in the comoving case because of the strong pushing effect of the moving interface.
Such modulation systems may be realized with optical pump-probe platforms~\cite{Ball_2025_knife} in the optical regime and switched transmission lines~\cite{Moussa_2023_observation} in the microwave regime.
Beyond its fundamental significance, the interluminal regime may enable new wave-based functionalities such as enhanced-diversity wave splitting and trapping, and provides a foundation for the study of outstanding problem in space-time systems, including acceleration~\cite{De_2025_scattering}, dispersion~\cite{De_2025_dispersion} and crystal diffraction~\cite{Deck_2019_uniform}.

\appendix

\section{Validations for Case II} \label{sec:Case_II}

\begin{figure*}[t]
    \centering
    \includegraphics[width=15cm]{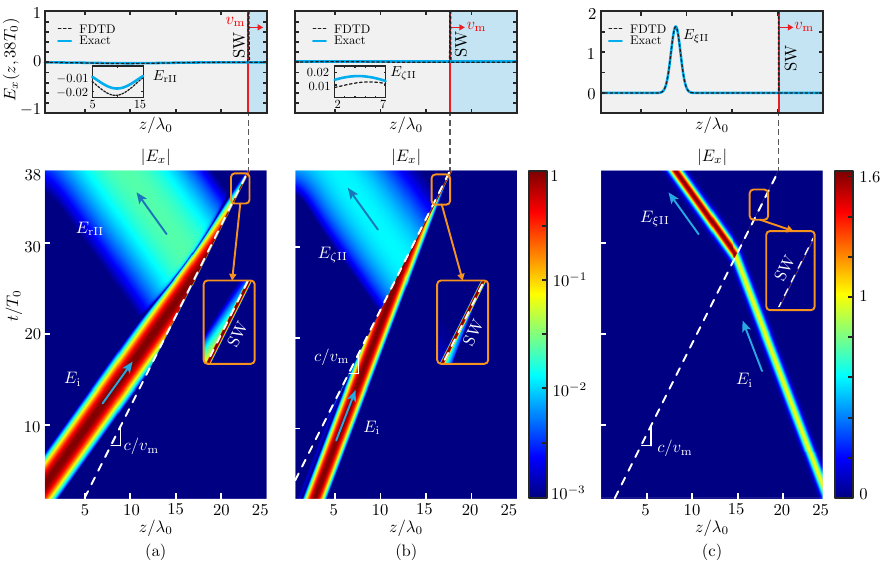}
    \caption{\label{fig:FDTD_CaseII} Full-wave (FDTD) simulations of a Gaussian pulse scattering at a comoving interface (Case II) with $v_{\m}=0.5c$. The media parameters are the same as in Fig.~\ref{fig:FDTD}. (a) Reflection case. (b) Later-backward-wave case. (c) Later-forward-wave case. The bottom panels show the electric-field magnitude $|E_x|$ in space-time diagrams normalized to $T_0$ and $\lambda_0$, with logarithmic scales used in (a) and (b) to highlight the small fields. The top panels show the corresponding fields at $t=38T_0$. `SW' denotes the shock wave.}
\end{figure*}

Figure~\ref{fig:FDTD_CaseII} presents the simulation results corresponding to Fig.~\ref{fig:FDTD} for Case II with $v_{\m}=0.5c$. Three incident cases, illustrated in the right panel of Fig.~\ref{fig:regimes}(c), give rise to the reflected wave in Fig.~\ref{fig:FDTD_CaseII}(a), the later-backward wave in Fig.~\ref{fig:FDTD_CaseII}(b), and the later-forward wave in Fig.~\ref{fig:FDTD_CaseII}(c). Shock waves are observed at the interface in all three cases, and their underlying mechanism is discussed in Sec.~\ref{sec:shock_waves}. 

%

\section{Energy-Momentum Relation in the Interluminal Regime} \label{sec:energy}

In~\cite{Li_2025_Energy}, we derived the energy-momentum relations between the wave and the modulation in the subluminal and superluminal regimes, while leaving the interluminal regime unresolved. In this section, we complete the theory using the results obtained in Sec.~\ref{sec:method}.

\begin{figure}[ht!]
    \centering
    \includegraphics[width=8.6cm]{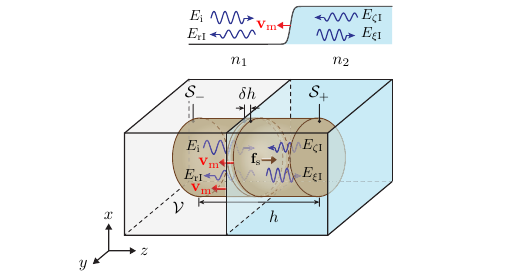}
    \caption{\label{fig:energy} Illustration of the imaginary cylinder comoving with the modulation interface in the contramoving interluminal regime. The top panel shows the corresponding moving step and the associated waves.}
\end{figure}

Integrating Maxwell’s equations in a comoving cylinder and taking the limit $h\rightarrow0$, as shown in Fig.~\ref{fig:energy}, provides the energy-momentum relations~\cite{Li_2025_Energy}
\begin{subequations} \label{eq:p_f}
        \begin{equation}\label{eq:p_vm}
        p_{\mathrm{s}}=\hat{\mathbf{z}} \cdot[\mathbf{S}]-v_{\m}[W]
    \end{equation}
    and
        \begin{equation}\label{eq:f_vm}
        \mathbf{f}_{\mathrm{s}}=\hat{\mathbf{z}} \cdot[\overline{\overline{\mathrm{T}}}]-v_{\m}[\mathbf{g}].
    \end{equation}
    where $[a]=a_{+}-a_{-}$ denotes the jump of $a$ across the interface. $p_{\mathrm{s}}$ and $\mathbf{f}_{\mathrm{s}}$ are the surface power and force densities at the interface. $W=(\mathbf{D}\cdot\mathbf{E}+\mathbf{B}\cdot\mathbf{H})/2$ is the energy density of the wave, $\mathbf{S}=\mathbf{E}\times\mathbf{H}$ is the Poynting vector, $\mathbf{g}=\mathbf{D} \times \mathbf{B}$ is the momentum density and $\overline{\overline{\mathrm{T}}}=\frac{1}{2}(\mathbf{D}\cdot\mathbf{E}+\mathbf{B}\cdot\mathbf{H})\overline{\overline{\mathrm{I}}}-\mathbf{D}\mathbf{E}-\mathbf{B}\mathbf{H}$ is the Maxwell stress tensor.
\end{subequations}
Furthermore, $p_{\mathrm{s}}$ and $\mathbf{f}_{\mathrm{s}}$ satisfy the power relation~\cite{Halliday_2013_fundamentals}
    \begin{equation} \label{eq:p_vf}
        p_{\mathrm{s}}=\mathbf{v_{\m}}\cdot\mathbf{f}_{\mathrm{s}}.
    \end{equation}

In Case~I, at the instant when the wave interacts with the interface, the scattered waves have the same phase $\phi$ as the incident wave. For simplicity, we consider the phase at its maximum and set the incident amplitude to~$1$. Under these assumptions, Eqs.~\eqref{eq:p_f} reduce to
\begin{subequations} \label{eq:ps_fs}
    \begin{equation}
    \begin{aligned}
        p_{\mathrm{s}}=\Bigg(\frac{\xi_{\mathrm{I}}^2}{\eta_2} - &\frac{\zeta_{\mathrm{I}}^2}{\eta_2} + \frac{r_{\mathrm{I}}^2}{\eta_1} - \frac{1}{\eta_1}\Bigg)\\ & - v_{\m} \Bigg( \frac{\xi_{\mathrm{I}}^2}{\eta_2 v_2 } +  \frac{\zeta_{\mathrm{I}}^2}{\eta_2 v_2 } - \frac{r_{\mathrm{I}}^2}{\eta_1 v_1} - \frac{1}{\eta_1 v_1}\Bigg)
  \end{aligned}
    \end{equation}
    and
    \begin{equation}
    \begin{aligned}
        f_{\mathrm{s}}=\Bigg(\frac{\xi_{\mathrm{I}}^{2}}{\eta_2 v_2}+&\frac{\zeta_{\mathrm{I}}^{2}}{\eta_2 v_2}-\frac{r_{\mathrm{I}}^{2}}{\eta_1 v_1}-\frac{1}{\eta_1 v_1}\Bigg)\\&-v_{\m}\Bigg(\frac{\xi_{\mathrm{I}}^{2}}{\eta_2 v_2^2}-\frac{\zeta_{\mathrm{I}}^{2}}{\eta_2 v_2^2}+\frac{r_{\mathrm{I}}^{2}}{\eta_1 v_1^2}-\frac{1}{\eta_1 v_1^2}\Bigg).
        \end{aligned}
    \end{equation}
\end{subequations}
Substituting the scattering coefficients [Eqs.~\eqref{eq:E_rI}-\eqref{eq:E_xiI}] into Eqs.~\eqref{eq:ps_fs} shows that the results satisfy the power-force relation~\eqref{eq:p_vf}, which indicates that the energy-momentum relations [Eqs.~\eqref{eq:p_f}] hold  in the contramoving interluminal regime.
However, these relations do not hold for Case II, where the formation of a shock wave at the interface (see Sec.~\ref{sec:shock_waves}) introduces singularities.

\begin{acknowledgements}
Z.L. acknowledges the helpful discussions and sharing with Dr. Zoé-Lise Deck-Léger. K.D.K. is supported by the Research Foundation -- Flanders (FWO) doctoral fellowship 1174526N.
\end{acknowledgements}  

\bibliography{Reference}

\end{document}